\documentclass[aps,prl,twocolumn,footinbib]{revtex4-1}
\usepackage{amssymb}
\usepackage{graphicx}
\usepackage{hyperref}
\usepackage{natbib}
\usepackage{dcolumn,amsmath,amsthm,amscd,amsfonts,amssymb,epsfig,graphics,graphicx,eucal}

 \hypersetup{pdfstartview={FitH},pdfpagemode={UseNone},
            colorlinks,linkcolor=red, citecolor=blue, urlcolor=blue,
            bookmarks=true, bookmarksopen=true, pdfnewwindow=true}

\usepackage{epstopdf}

\newcommand{\rmi}{{\rm i}}
\newcommand{\rmd}{{\rm d}}

\renewcommand {\Re}{\mathop{\mathrm{Re}}\nolimits}

\begin{document}

\title{Observation and control of nonlinear electromagnetic topological edge states}

\author{\firstname{D.~A.} \surname{Dobrykh$^{1}$},
  \firstname{A.~V.} \surname{Yulin$^{1}$},
  \firstname{A.~P.} \surname{Slobozhanyuk$^{1,2}$},
  \firstname{A.~N.} \surname{Poddubny$^{2,1,3}$},
    \firstname{Yu.~S.} \surname{Kivshar$^{1,2}$}}

\affiliation{$^1$ITMO University, St. Petersburg 197101, Russia\\
$^2$Nonlinear Physics Center, Australian National University, Canberra ACT 2601,
Australia\\
$^3$Ioffe Physical-Technical Institute of the Russian Academy of Sciences,
St. Petersburg 194021, Russia}


\begin{abstract}
Topological photonics has recently emerged as a route to realize robust optical circuitry,
and nonlinear effects are expected to  enable tunability of topological states with the light intensity.
Here we realize experimentally nonlinear self-induced spectral tuning of the electromagnetic topological edge states
in an array of coupled nonlinear resonators in a pump-probe experiment. In a weakly nonlinear regime,
we observe that  resonators frequencies exhibit spectral shifts, that are concentrated mainly at the edge mode affecting only weakly the bulk modes. For a strong pumping, we describe several scenarios of the transformation of the edge states
and their hybridization with bulk modes, and also predict a parametrically driven transition from topological to unstable regimes.
\end{abstract}


\maketitle

{\it Introduction.} Topological photonics has recently emerged as an universal tool to achieve disorder-immune guiding of electromagnetic waves~\cite{Lu2014,Lu2016,Khanikaev2017,Ozawa2018}, but practical realizarions of topological photonic devices require dynamic tunability. Despite the latest experiments on topological lasers~\cite{StJean2017,Barik2018,Segev2018b,Iorsh2018} and several theoretical proposals for tunable nonlinear devices~\cite{Zhou2017,Kartashov2017,Solnyshkov2018,Savelev2018}, experimental demonstrations of reconfigurable topological photonic structures are still missing. An important recent milestone is the demonstration of nonlinearity-driven topological transition in electric circuits~\cite{Hadad2018}. However, to the best of our knowledge, nonlinear tunability of topological electromagnetic edge states has not been shown so far.

\begin{figure}[b!]
\protect\includegraphics[width=0.9\columnwidth]{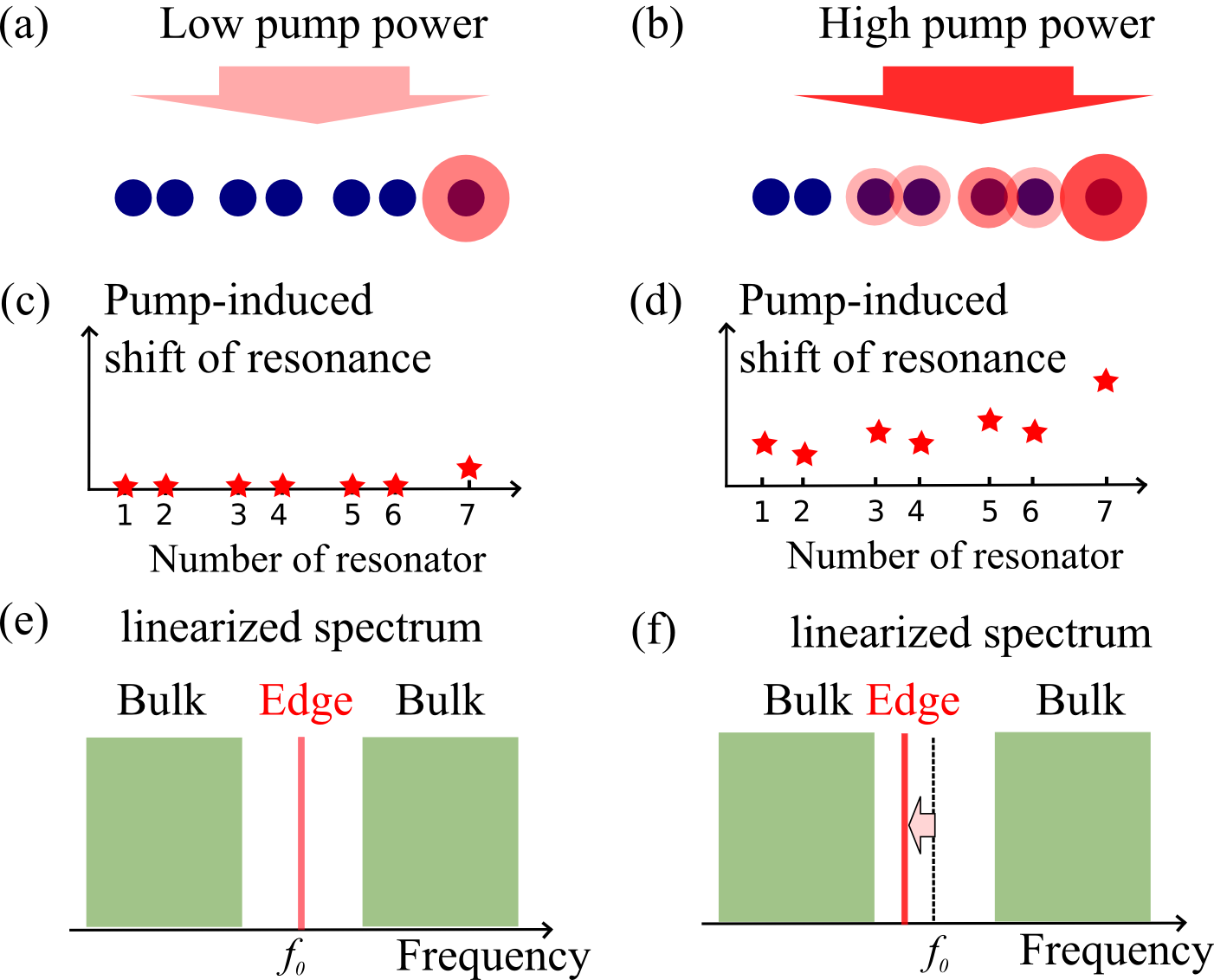}
\caption{(a,b) Arrays of resonators under (a,c,e) weak and (b,d,f) strong external homogeneous pumps.
The pump is show by red colors. (c,d) Pump-induced spectral shifts of the resonant frequencies.
(e,f) A change the linear spectrum of both bulk (green) and edge (red) states.
A dashed vertical line marks the resonant frequency in the linear regime. \label{fig:1}}
\end{figure}

In this Letter, we realize self-induced nonlinear spectral tuning of topological edge states of electromagnetic waves in one-dimensional arrays of identical electromagnetic resonators with Kerr-type nonlinearity. The resonators are separated by alternating long and short links, being coupled electromagnetically. The concept of nonlinearity-induced topological tuning is illustrated in Fig.~\ref{fig:1}. In the linear regime (a), an array of resonators is described by the Su-Schriffer-Heeger (SSH) model, and it represents a simple one-dimensional topological structure~\cite{Shen2013}. The system supports an electromagnetic topological mode localized at the edge terminated by the long link (the right edge for the structure shown in Fig.~\ref{fig:1}).  This state occurs at the resonant frequency of an individual resonator being localized spectrally in the center of the band gap that appears due to a difference of the coupling coefficients for long and short links of the SSH model. When the structure is pumped homogeneously at the central frequency  (see Fig.~\ref{fig:1}, left column), nonlinear spectral shifts of the resonator frequencies are induced. These spectral shifts modify the spectrum of linearized excitations of the structure, as shown in the bottom row of Fig.~\ref{fig:1}. Since the pump is at the resonance, the spectral shift is at its maximum for the edge resonator,  affecting only  the corresponding  edge state while the bulk linearized spectrum remains mainly unchanged~{(Fig.~\ref{fig:1}f)}. As such, even though the pump is homogeneous, resonators are identical, and all tunneling coupling links are pump-independent (contrary to the concept of Refs.~\cite{Hadad2018,Hadad2016}), the edge state exhibits a nonlinear spectral shift. At stronger  pumping we expect that the edge state is destroyed due its interaction with the bulk modes, while the nonlinear spectral shifts become large and identical for all the resonators.

This description grasps our concept of a nonlinear control of topological edge states but the actual physics of nonlinear topological systems is considerably richer~\cite{Leykam2016,Solnyshkov2016,Nori2017,Kartashov2017}. Specifically, one has to take into account the intensity dependence of the stationary nonlinear distribution of the electromagnetic fields at the pump frequency.
Another important effect is the parametric nonlinear interaction between bulk and edge modes in the linearized system. An interplay of linear localization with nonlinearity and the fate of the edge state at high pumping is a complex problem.
Next, we present a rigorous theoretical model accounting for these effects and then  discuss  experimental results in the pump-probe setup.


{\it A topological chain of nonlinear oscillators.}
To introduce the concept of nonlinear topological states, we consider a topological array of coupled
oscillators with cubic nonlinearity described by a system of equations
\begin{equation}
 \frac{\rmd a_{n}}{\rmd t}=-\gamma a_{n}-\rmi|a_{n}|^{2}a_{n}+t_{n,-}a_{n-1}+
 t_{n,+}a_{n+1}+ P  \:,\label{eq:main}
\end{equation}
where $a_{n}$ is a normalized amplitude of the $n$-th oscillator ($n=1\ldots N$), $\gamma$ is a damping  coefficient,
and $P$ is an amplitude of resonant homogeneous pump. Alternating strong and weak nearest-neighbor couplings are introduced as
$t_{0,-}=t_{N,+}=0$, $t_{2k,-}=t_{2k-1,+}=t_{1}$, $t_{2k,+}=t_{2k-1,-}=t_{2}$. Equation~\eqref{eq:main}
is rather general, and it can be applied to different systems~\cite{Chen2014, Solnyshkov2017}. In the case of electromagnetic
resonators, $a_{n}$ stand for the slow-varying complex amplitudes of the current.
Our goal is to find the stationary states excited by the pump and study the linear spectrum and stability.

We distinguish here two fundamentally different problems. In a general case, both nonlinear stationary states and linearized modes excited
on the background of nonlinear solutions can be either localized or delocalized. When the pump is at the resonance, we expect
a stationary localized nonlinear mode to exist. On the other hand, the spectrum of linear excitations can include localized modes (such as edge states)
as well. However, in those cases the corresponding pump dependences differ for nonlinear stationary modes
and linearized edge states.

\begin{figure}[t!]
\protect\includegraphics[width=1.0\columnwidth]{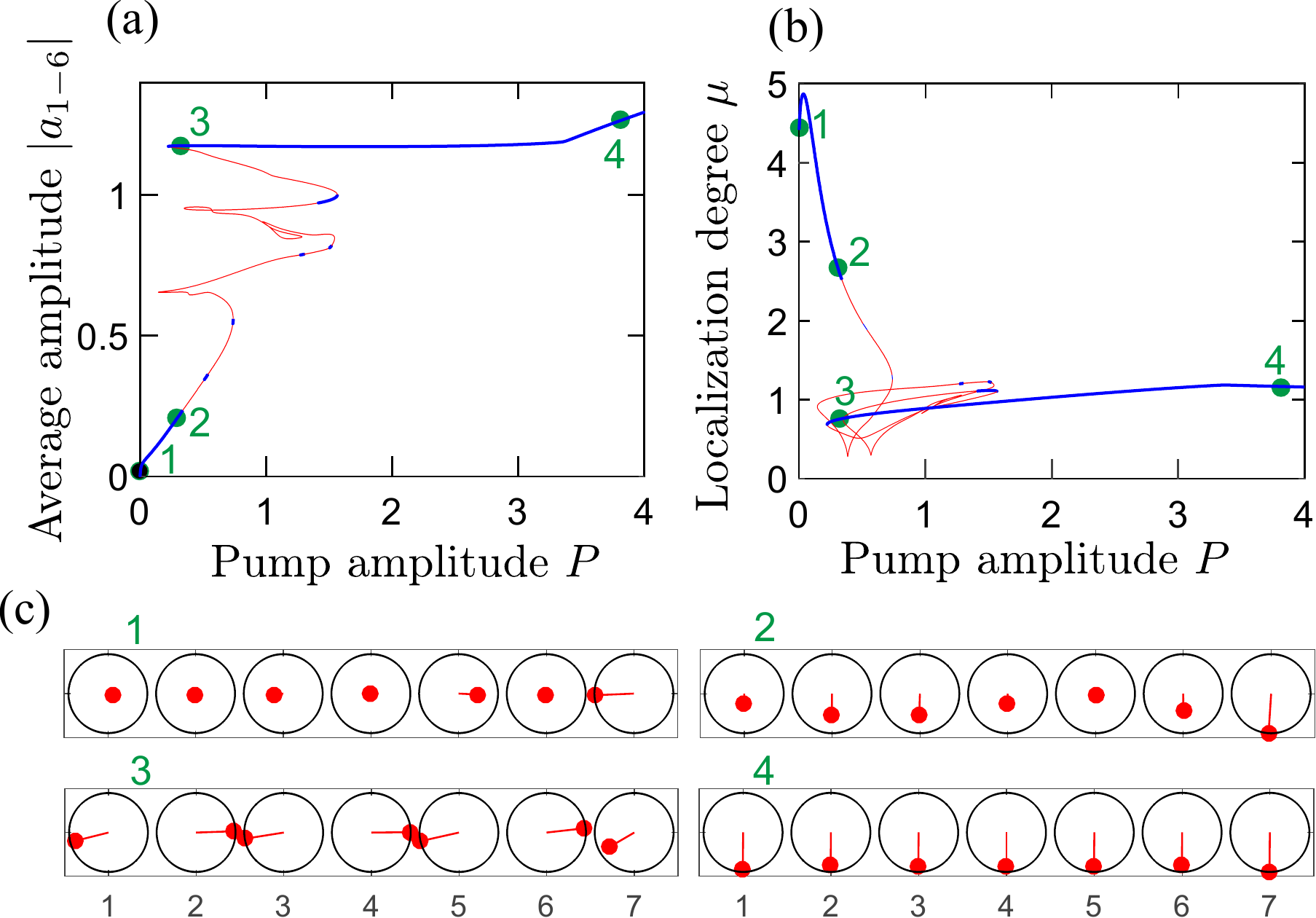}
\caption{ (a) Average resonator amplitude  $|a_{1-6}|$ and (b) localization degree $\mu$ for the amplitudes of the bulk and edge states vs. pump amplitude.
Blue (red) lines correspond to linearly stable (unstable) solutions. Panel (c) shows  spatial distributions of the normalized amplitudes
 at the points 1-4 on the bifurcation diagram. Distance of a red spot from the center of the ring is proportional to the absolute
 value of the amplitude, whereas the angle shows its phase  with respect to the pump with  in-phase oscillations corresponding
 to the  direction to the left. Calculation parameters are $\gamma=0.02$, $t_{1}=1$, $t_{2}=0.48$.
 \label{fig:theory-stationary}}
\end{figure}

First, we consider high-quality resonators assuming $\gamma\ll |t_{1,2}|$.
We study the chain of  $N=7$ resonators with  the right-most coupling  being $|t_{2}|<|t_{1}|$ so that a localized topological
state is formed at the right edge ($n=N$). The results for stationary states are summarized in Fig.~\ref{fig:theory-stationary}. Panel (a) shows the average amplitude in the bulk, defined as $|a_{1-6}|^{2}=\sum_{n=1}^6 |a_n|^2/6$, vs. the pump amplitude $P$.
For $1\lesssim P\lesssim 2$, these results suggest that our system exhibits bistability (red lines).

\begin{figure}[b!]
\includegraphics[width=1.\columnwidth]{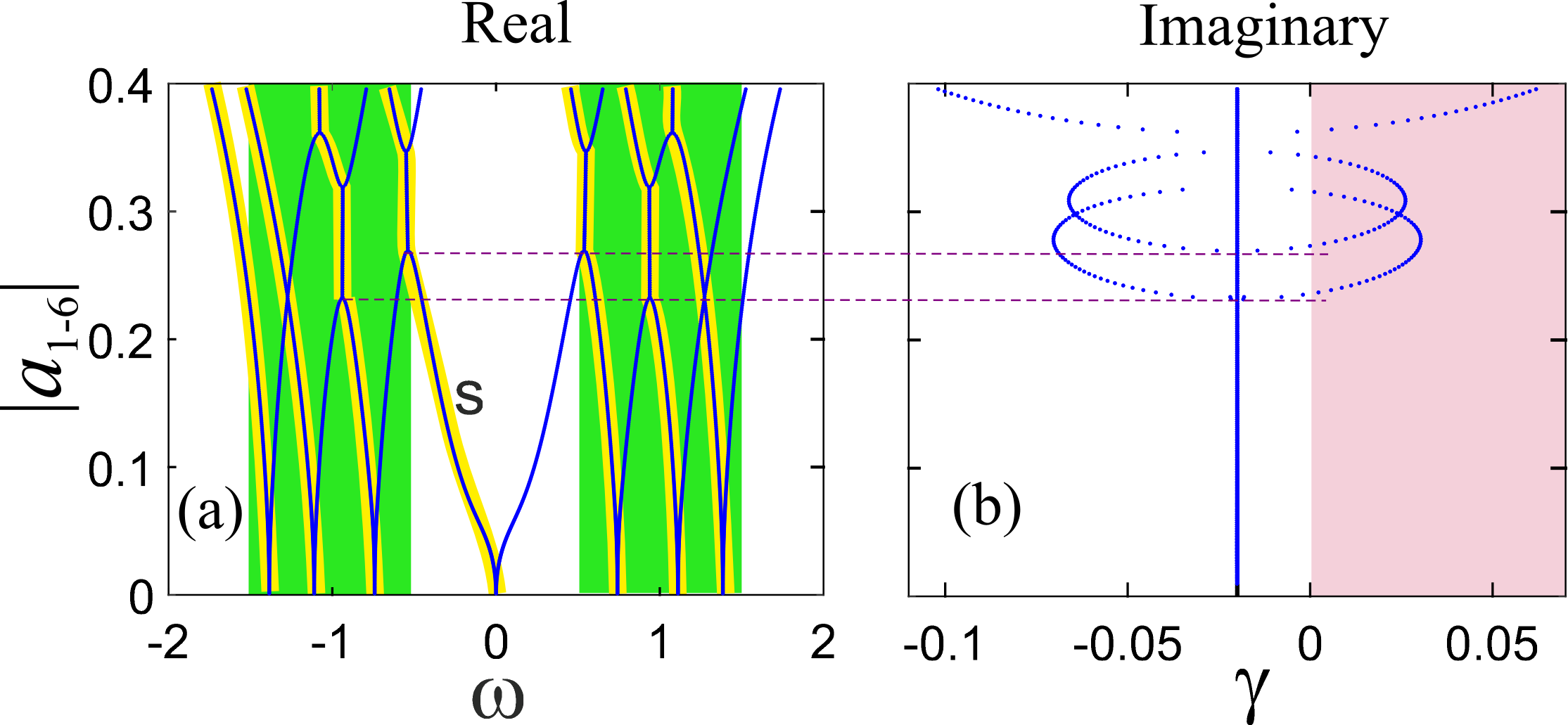}
\caption{Real (a) and imaginary (b) parts of the numerically calculated eigenfrequencies of the modes as functions of the average amplitude $|a_{1-6}|=\sqrt{\sum_{n=1}^6 |a_n|^2 /6}$. The eigenfrequencies with positive imaginary part [red area in panel (b)] correspond to the unstable modes.  The yellow stripes show the approximate positions of the eigenvalues if the parametric effects are disregarded.  The edge state  is denoted as ``$s$''. The green rectangles show the frequencies of the continuum of the linear excitations on zero background.
The calculation parameters are the same as in Fig.~\ref{fig:theory-stationary}.
}\label{fig:theory-spectrum}
\end{figure}

Our main goal is to study the edge states, so we examine a difference of the  amplitude of the last oscillator and the amplitudes in the bulk oscillators. A localization strength can be defined by the edge-to-bulk ratio $\mu \equiv |a_{1-6}|/|a_7|$ shown in Fig.~\ref{fig:theory-spectrum}(b) as a function of the pump. We observe that at low values of the pump, the localization
parameter $\mu$ is considerable larger than unity indicating the presence of a stationary edge state. Indeed,
from  the spatial distribution of the stationary state [Fig.~\ref{fig:theory-spectrum}(c)] we observe that at low pump values the amplitude is the largest for  the last resonator.   For the chosen parameters, the stationary edge state survives until the whole system looses its stability at a threshold pump value, as shown in  Fig.~\ref{fig:theory-spectrum}(c) for the points $1$ and $2$ of the bifurcation diagram. For larger values of the pump, the edge state is eventually destroyed. Correspondingly, the localization  parameter $\mu$ approaches the unity,
and the spatial  distributions for the points $3$ and $4$  in Fig.~\ref{fig:theory-spectrum}(c) become homogeneous.

Eigenvalues of the linear spectra on the background of nonlinear modes are shown in Fig.~\ref{fig:theory-spectrum} as functions of
the average intensity of six resonators. For weak intensities, the spectrum can be separated into the bulk modes lying inside the allowed bands of the infinite structure
($\omega\in \pm [t_{1}-t_{2}\ldots t_{2}+t_{2}]$, green rectangles in Fig.~\ref{fig:theory-spectrum}(b))
 and the edge mode ($\omega=0$).  Importantly, the linearization produces a parametric term $a_{n}^{2}\delta a_{n}^{*}$, where $\delta a_{n}$ is the excitation amplitude. Hence, the linear modes contain two parametrically coupled harmonics (cf. the Bogoliubov topological polaritonic modes in Ref.~\cite{Liew2016}). That is why the eigenfrequencies are double degenerate in the linear limit but at a finite pump the degeneracy is lifted. The yellow stripes mark the eigenvalues which are close to the eigenvalues of the system with omitted parametric term, and they have much larger excitation efficiency. In agreement with our expectations in Fig.~\ref{fig:1}, the central yellow stripe corresponding to the edge state (marked by ``s'') exhibits a nonlinear shift, much stronger than that for bulk modes.

If the pump  becomes strong enough, the instability sets in, as seen from the imaginary parts of the eigenfrequencies in Fig.~\ref{fig:theory-spectrum}(b). At low intensities, the eigenfrequencies are purely real, then some of the eigenvalues collide producing pairs of complex conjugated eigenvalues. When the real part of an eigenvalue becomes positive (i.e., it enters the red region) the system acquires an exponentially growing perturbation and becomes unstable. Depending on parameters, the first  instability can be produced by a collision of the edge state with a bulk mode or by a collision of two bulk modes. Summarizing the results in Figs.~\ref{fig:theory-stationary} and \ref{fig:theory-spectrum}, we can claim that the mechanism of the  edge state collapse is a dynamical instability developing in the system. The nonlinear stage of instability is studied by direct numerical simulations, and it is found that the system undergoes a complex dynamics.

However, the pump-dependent dependence described above is not the only scenario. In the case of higher losses,
$\gamma\gtrsim |t_{1,2}|$, the parametric instability becomes less important. In this case, the localization length of the edge state grows monotonically with the pump due to hybridization with the bulk modes until it gets eventually delocalized while the system remains stable. The case of high losses is analyzed  in detail in Supplementary Materials\footnote{See online Supplementary Materials}, see Fig.~S1 and Fig.~S2.

\begin{figure}[b!]
\protect\includegraphics[width=1.0\columnwidth]{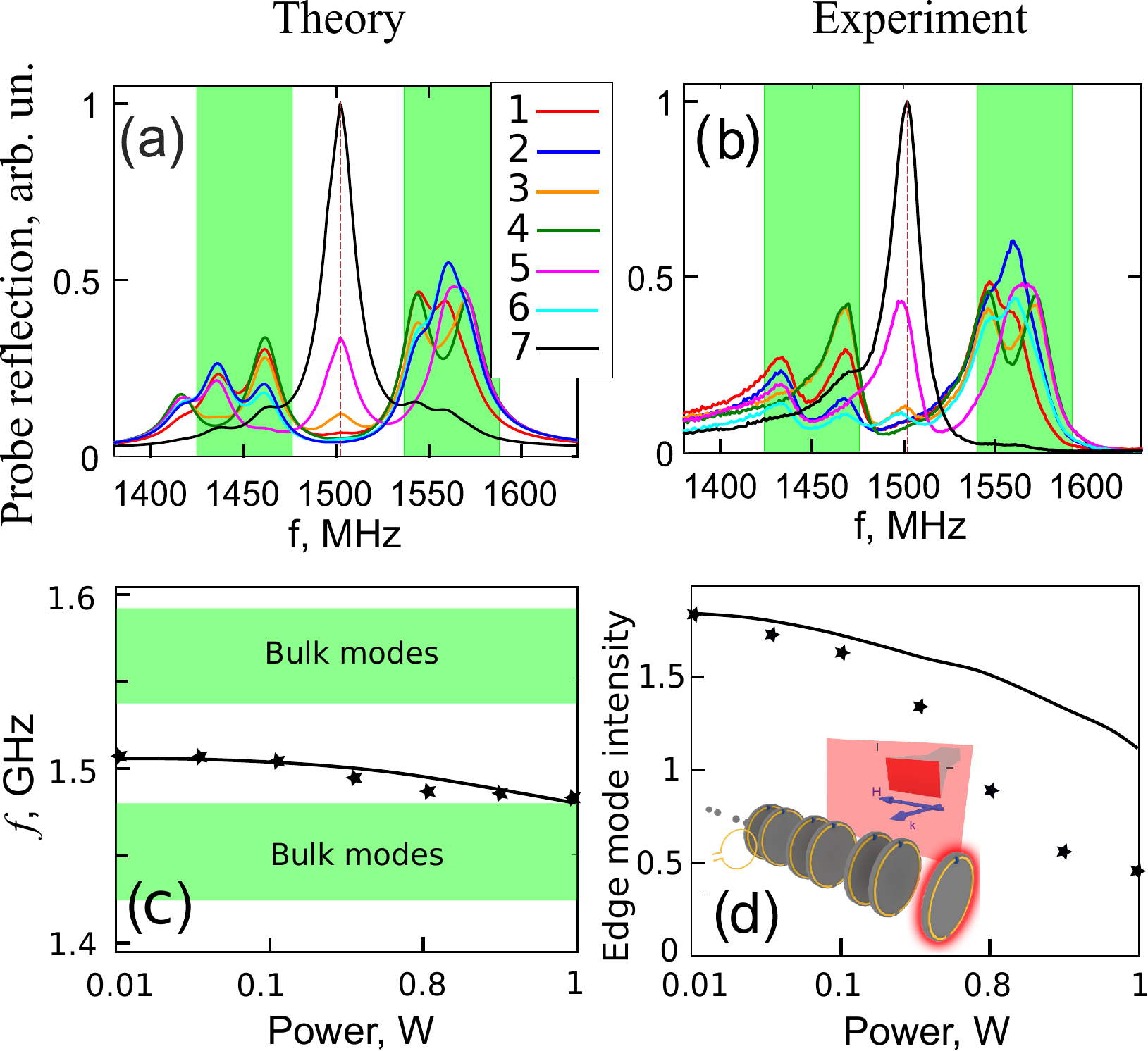}
\caption{(a,b) Spatially-resolved pump-probe spectra calculated numerically and measured experimentally.
(c) Frequency of the edge state vs. pump power. A black solid line corresponds to theoretical results, and dots stand
for experimental data. Green regions mark the allowed bands of an infinite structure calculated in the nearest-neighbor approximation.
(d) A ratio of the amplitude of the spectral maximum associated with the edge state and the amplitude of the most intensive spectral maximum associated with a bulk mode. A blue solid  line corresponds to numerical results and dots mark experimental data. Curves  1--7 correspond to  the resonators from left to right.
} \label{fig:exp}
\end{figure}

{\it Pump-probe experiment.} Our experimental results are summarized in Fig.~\ref{fig:exp}.
In experiment, we have fabricated an array of $N=7$ broadside-coupled split-ring resonators with the magnetic dipole  resonance at the $f_{0}\approx 1500~$MHz frequency. A varactor diode has been mounted inside the gap of each ring to provide the nonlinear tunability of the frequency~\cite{Filonov2016}. More details are given in \cite{Note1}.
The spectrum of linearized excitations, theoretically considered in Fig.~\ref{fig:theory-spectrum}, can be directly accessed in the spatially-resolved pump-probe setup. The experimental scheme is sketched in the  inset of Fig.~\ref{fig:exp}(d). The monochromatic homogeneous pump at the  resonance $f=f_{0}$ has been provided by a
rectangular horn antenna. The probe  signal has been measured near each resonator by a small
loop antenna connected to the receiving port of the analyzer.  The probe spectrum of the  reflection coefficient $\Re S_{11}$ for the loop antenna has been determined as  a function of its position and  the  pump power.

Figures~\ref{fig:exp}(a, b) show both theoretical and experimental spectra in the linear regime.
 A spectral signature of the edge state is clearly seen by the presence of  the central resonance at the frequency $f_{0}$,  when the probe is close to the right edge (black and magenta curves for $n=5, 7$, respectively). Interestingly, the spectrum at the second-to-last resonator, $n=6$ (cyan curve) does not manifest a central resonance. This reflects the spatial structure of the edge state in the Su-Schrieffer-Heeger model, that, in the nearest-neighbor approximation, has nonzero amplitudes only at the modes with odd $n$.
 Other resonances in the spectrum correspond to bulk modes, and they are  spectrally located within the allowed bands of the infinite array (shaded green rectangles). The coupling coefficients between the resonators are fitted to match the experimental spectrum with the initial values provided by the rigorous full-wave simulations in the CST Microwave Studio package. We also include  the coupling of the resonators to the next neighbors which is taken to be $\approx 17$ times weaker then the coupling in dimers, $t_{3}=0.06t_{1}$, with $t_{1}=55$~MHz, $t_{2}=0.48t_{1}$. The experimental pump power  is related to the dimensionless amplitude in Eq.~\eqref{eq:main} as $P_{\rm exp}({\rm W})=0.43 \cdot P^2 $.  More details are given in \cite{Note1}. We take into account weak nonlocality of the probe, that senses the current not only in the nearest resonator but also in the adjacent resonators. Namely, the signal measured at $n$-th resonator is given by
$E_n=a_n+\chi_1 a_{n+1}+\chi_2 a_{n-1}+\chi_3(a_{n+2}+a_{n-2})$ for odd $n$, and $E_n=a_n+\chi_1 a_{n-1}+\chi_2 a_{n+1}+\chi_3(a_{n+2}+a_{n-2})$, for even $n$, where the coupling coefficients are $\chi_{1}=-0.12$, $\chi_{2}=-0.1$ and $\chi_{3}=-0.075$.  A negative sign comes from anisotropy of the magnetic dipole. Nonlocality results in asymmetry of the excitation efficiencies for upper and lower allowed bands. Since the upper band corresponds to odd Bloch functions and $\chi_{1}<0$, it is sensed by a probe with higher efficiency than for the bottom band.

Our main experimental result is summarized in Fig.~\ref{fig:exp}(c) for the edge state frequencies extracted from the probe spectra as a function of the pump power (raw spectra are given in Fig.~S3 \cite{Note1}). The central peak, corresponding to the edge mode, shifts with the pump power approaching the bulk bands. Experimental data (stars) are in a good agreement with our theory (black curve). This provides a direct experimental observation of the nonlinearly tunable electromagnetic topological edge state. We also compare the amplitude of the spectral maximum corresponding to the edge state to the most intense spectral maximum associated with a bulk mode. This ratio is plotted in Fig.~\ref{fig:exp}(d) as a function of the pump power (stars). The edge state becomes less pronounced for higher pump intensities.

\begin{figure}[t!]
\includegraphics[width=1.0\columnwidth]{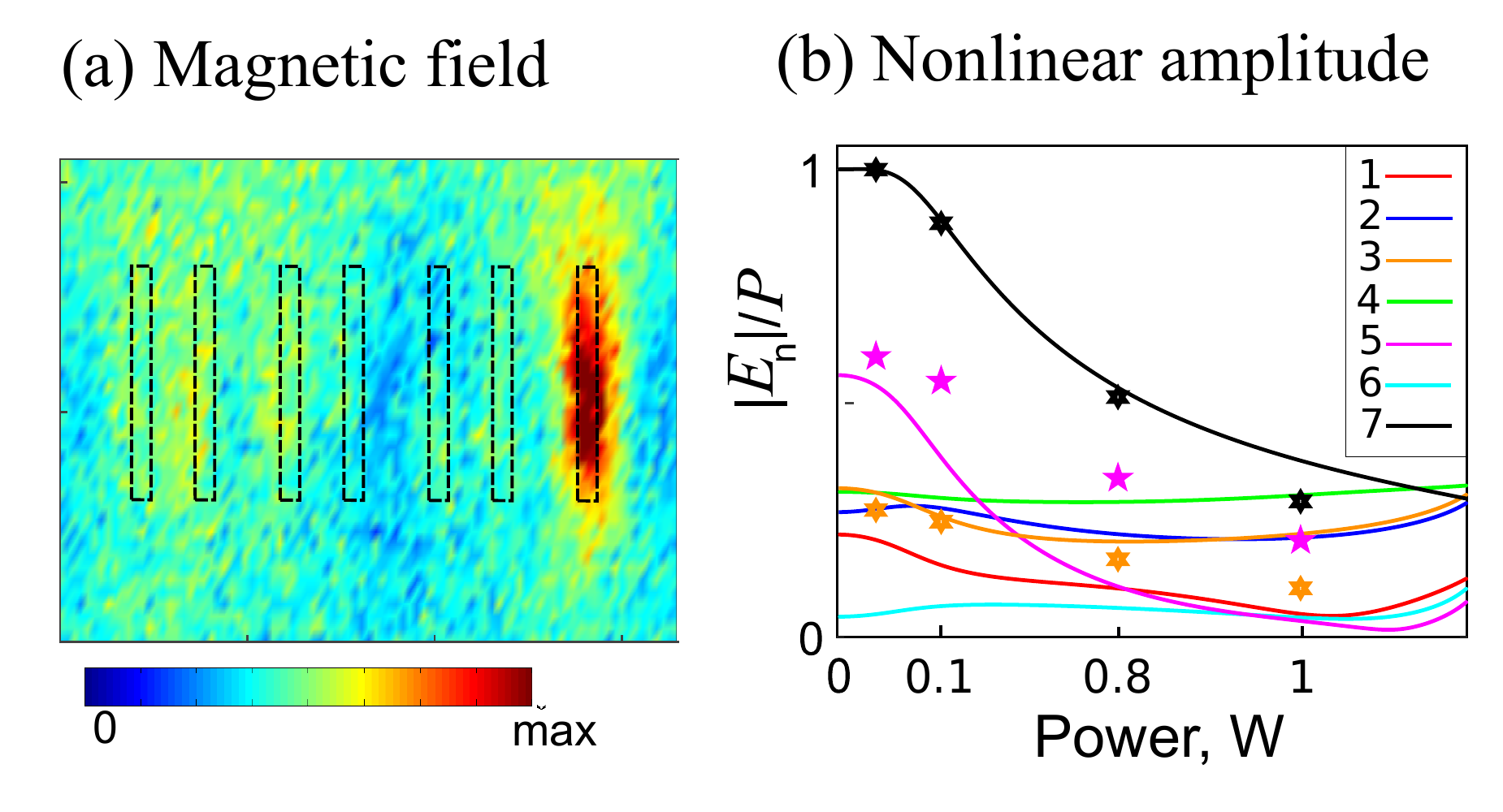}
\caption{ (a) Magnetic field measured at the distance $\l=2$ mm from the structure. (b) A ratio of the normalized amplitudes of the bulk and edge resonators vs.  pump power. Solid lines correspond to numerical simulations, and dots mark the experimental data.}\label{fig:exp-stationary}
\end{figure}

A  decay of the edge state is trustworthy established fact but the specific mechanism of this decay is not evident from experiments. The absence of the discontinuities (a part of the curve with a steep slope) at the dependencies of the experimentally measured oscillation amplitudes on the pump power suggests that the instability destroying the edge state does not happen in experiment. At low pump power, we observe a good agreement between the theory and experiment.  The development of instability at high intensities is a more subtle issue depending on dispersion and nonlinearity of the resonators. The precision of the coupling coefficients fitting needed to reproduce and to explain the oscillation dynamics in the linear regime is considerably lower than that needed to describe accurately the onset of the instability. It is also worth noticing that in real experiment the parameters of the split ring resonators are not identical. First, the antenna used to excite and to measure the oscillations in the resonators introduces an inhomogeneity not accounted by the model. Secondly, the variation of the resonator parameters  is caused not only by the inaccuracies of the manufacturing of the electrodes but by the variation of the parameters of the diodes used in the experiment. This possibly explains why theoretical and experimental results match better in the linear regime.
Moreover,  the theory does not fully account for the actual nonlinearity of the resonators.  One can consider nonlinear damping effect i.e. a dissipative nonlinearity, see Supplementary Fig.~S4 in \cite{Note1} . Another possible effect is the decrease of the oscillator coupling to the pump at high pump strength. The analysis of such effects is presented in Fig.~S5 in \cite{Note1} and it shows, that their consideration can significantly improve the  agreement between theory and experiment. Further improvements require accurate account for real nonlinearity which is saturable and possibly not instantaneous. It is also possible that the long-range coupling becomes more important in the nonlinear case. The development of a quantitatively  precise description is out of the scope of the present work. The main experimental result, nonlinear frequency tuning of edge state in Fig.~\ref{fig:exp}(c) is well described by a simplistic model.

We also analyze a stationary map of the electromagnetic field at the pump frequency vs. pump power. Instead of measuring reflection $S_{11}$ for an emitting probe loop antenna, we measure transmission coefficient $S_{12}$ from a horn antenna (a pump) to a probe loop antenna. The input signal at the frequency $f_{0}$ is sent to the pump antenna only.  Figure~\ref{fig:exp-stationary}(a) shows the measured data for the magnetic field for low values of the pump. The field is localized at the edge resonator providing a direct experimental evidence of the edge state. When the pump power increases, the stationary field becomes delocalized, as shown in Fig.~\ref{fig:exp-stationary}(b). Blue (black) stars  show the power dependencies for a ratio of the measured signal $E$ to the pump power for the resonators with $n=5$ and $n=7$.  In the same panel, we show the calculated dependencies of the same ratio $E/P$.  The calculated amplitude for $n=7$ matches experimental data reasonably well. The signal measured from the oscillator $n=5$ has the second largest amplitude, and being relatively well described by our theory. For large pump power, the stationary state becomes delocalized. A background of the signal, corresponding to $n<5$, depends weakly on the pump power, being governed likely by long-range couplings.

{\em Conclusion.} We have studied nonlinear self-induced tuning of the electromagnetic topological edge states in arrays of coupled nonlinear resonators with alternating weak and strong couplings described by the nonlinear topological SSH model. Combining nonlinear transmission and pump-probe setups,  we have demonstrated experimentally both stationary nonlinear states and linearized modes. Frequency tuning of the edge states has been observed being in a good   agreement with theoretical calculations.  We have revealed different scenarios of the pump-induced decay of topological edge states. Our results provide important insights into the physics of nonlinear tunable topological structures.

\begin{acknowledgments}
We acknowledge useful discussions with A. Khanikaev, D. Powell,  and I. Shadrivov. This work was supported by the Russian Foundation for Basic Research (grant 15-32-20866) and the Australian Research Council. ANP acknowledges a partial support of the ``Basis'' Foundation. The work of AVY was supported by the Government of the Russian Federation (grant 074-U01) through the ITMO Fellowship.
\end{acknowledgments}


%

\end{document}